\documentclass[10pt]{IEEEtran}
\usepackage[final]{graphicx}
\usepackage{subfigure}
\graphicspath{{./}{./figures/}}

\newcommand{\ring}{{\varphi}}
\newcommand{\tran}{{T}}
\newcommand{\g}{{G}}

\newcommand{\sa}{{\cal A}}
\newcommand{\si}{{s}}

\newcommand{\saa}{{\cal B}}

\newcommand{\darkbox}{\vrule height6pt width6pt depth0pt}

\newcommand{\comment}[1]{ }
\newtheorem{theorem}{{{\bf Theorem}}}

\newtheorem{proposition}{{ {\bf Proposition}}}
\newtheorem{definition}{{ {\bf Definition}}}
\newtheorem{lemma}{{{\bf Lemma}}}

\title{A Simple Algorithm for Coloring m-Clique Holes}

\author{Bechir Hamdaoui \\~\\ {\small School of Electrical Engineering and Computer Science~\\ Oregon State University, Corvallis, OR 97331 \\ \small Email: hamdaoui@eecs.oregonstate.edu.}}

\begin{document}

\maketitle

\begin{abstract}
An m-clique hole is a sequence $\phi=(\Phi_1,\Phi_2,\dots,\Phi_m)$ of $m$ distinct cliques such that $|\Phi_i| \leq m$ for all $i=1,2,\ldots,m$, and whose clique graph is a hole on $m$ vertices. That is, $\phi$ is an m-clique hole if for all $i\neq j$, $i,j=1,2,\ldots,m$, $\Phi_i \cap \Phi_{j} \neq \emptyset$ if and only if $(j-1)~\mbox{mod}~m = (j+1)~\mbox{mod}~m = i~\mbox{mod}~m$. This paper derives a sufficient and necessary condition on m-colorability of m-clique holes, and proposes a coloring algorithm that colors m-clique holes with exactly m colors. 
\end{abstract}


\section{Introduction}
\label{sec:int}
The intersection graph of a family of non-empty sets is the graph whose vertex set is the sets of the family, and whose edge set is all unordered pairs of vertices whose corresponding sets in the family intersect. A clique in a graph is a complete subgraph maximal under inclusion. The clique graph K(\g) of a graph \g~is the intersection graph of the cliques of \g.

A hole is a chordless cycle on at least four vertices. A hole is odd if it has an odd number of vertices.
Consider a sequence $\phi=(\Phi_1,\Phi_2,\dots,\Phi_m)$ of $m$ distinct cliques such that $|\Phi_i| \leq m$ for all $i=1,2,\ldots,m$. We call the sequence $\phi$ an {\em m-clique hole} if the clique graph K($\phi$) is a hole on $m$ vertices; i.e., if for all $i\neq j$, $i,j=1,2,\ldots,m$, $\Phi_i \cap \Phi_{j} \neq \emptyset$ if and only if $(j-1)~\mbox{mod}~m = (j+1)~\mbox{mod}~m = i~\mbox{mod}~m$.
We call $\phi$ an odd m-clique hole if K($\phi$) is an odd hole. Otherwise, $\phi$ is an even m-clique hole.

An m-clique hole $\phi$ is said to be m-colorable if one can color
all the vertices in $\phi$ with m different colors such that no two vertices in $\phi$ sharing an edge between them are colored with the same color. An odd m-clique hole is imperfect~\cite{crst:06},
and hence its chromatic number is greater than its clique number $m$ for some of its induced subgraphs~\cite{berge:61}; i.e., odd m-clique holes are not m-colorable in general.
In this paper, we first provide and prove a sufficient and necessary condition on m-colorability of m-clique holes, and then provide a coloring algorithm that colors m-clique holes with exactly m different colors. 

\section{A Sufficient and Necessary Condition}
\label{sec:theorem}
In this section, we present and prove the main theorem of the paper
which states a sufficient and necessary condition on the
colorability of m-clique holes. The proof will also be the basis for the proposed coloring algorithm that we present in Section~\ref{sec:algorithm}.
As we go through the proof, we will introduce and prove several lemmas and propositions that will be used for proving the main theorem as well as for designing the proposed coloring algorithm. The proof will go until the end of this section. Throughout this paper, for simplicity of notation, all indices wrap around after reaching m; e.g., $\Phi_{m+1}$ refers to $\Phi_{1}$, $\Phi_{m+2}$ refers to $\Phi_{2}$, etc., and $\Phi_{0}$ refers to $\Phi_{m}$, $\Phi_{-1}$ refers to $\Phi_{m-1}$, etc. Now, we state the main result.

\begin{theorem}\label{theorem:1}
An m-clique hole $\phi=(\Phi_1,\Phi_2,\dots,\Phi_m)$ is m-colorable
if and only if
$\sum_{i=1}^{m}{|\Phi_i \cap \Phi_{i+1}|} \leq m{\cal
b}\frac{m}{2}{\cal c}$.
\end{theorem}

\subsection{The "if" part proof}
Let us denote the set of
vertices $\Phi_i \cap \Phi_{i+1}$ by $\sa_i$ for all
$i=1,2,\ldots,m$. Suppose that $|\Phi_i| \leq m$ for all
$i=1,2,\ldots,m$, and $\sum_{i=1}^{m}{|\sa_i|} \leq m{\cal
b}\frac{m}{2}{\cal c}$. In this section, we will show that $\phi=(\Phi_1,\Phi_2,\dots,\Phi_m)$ has a proper
m-coloring by proposing an algorithm that colors $\phi$ with exactly m different colors.

Note that any vertex in $\Phi_i - \{\sa_{i-1} \cup \sa_{i}\}$
shares edges with and only with vertices in $\sa_{i-1} \cup
\sa_i$. Now since $|\Phi_i|\leq m$ and $\{\sa_{i-1} \cup \sa_i\}
\subseteq \Phi_i$, then the vertices in $\Phi_i - \{\sa_{i-1} \cup
\sa_{i}\}$ can clearly be m-colored provided that the vertices in
$\sa_{i-1} \cup \sa_i$ already have a proper m-coloring.
Therefore, in order for us to color $\phi$ with $m$ colors, it
suffices to provide the vertices in $\cup_{i=1}^{m}\sa_i$ with a
proper m-coloring. In what follows, we will be concerned only with
coloring the vertices in $\cup_{i=1}^{m}\sa_i$.

\begin{figure}
  \centerline{
  \includegraphics[width=0.75\columnwidth]{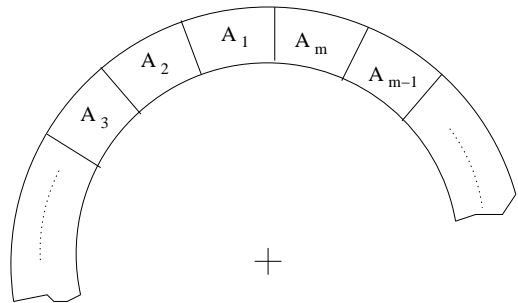}}
  \caption{Ring representation of $\cup_{i=1}^{m}\sa_i$.}\label{fig:ring}
\end{figure}

For convenience, let us represent the induced graph from $\phi$
consisting of only vertices in $\cup_{i=1}^{m}\sa_i$ as a ring
$\ring$ as shown in Fig.~\ref{fig:ring}.
In this representation, each sector of the ring corresponds to a
set $\sa_i$. Observe that each vertex in $\sa_i$ shares an edge
with and only with every vertex in its own sector $\sa_i$ and its
two immediate adjacent sectors $\sa_{i-1}$ and $\sa_{i+1}$.
Also, note that $\sa_i \cap \sa_j=
\emptyset$ for $i\neq j$, since otherwise
$\Phi_{i'}\cap \Phi_{j'} \neq \emptyset$ holds for some $j'\neq
i'+1$; which contradicts the definition of m-clique holes.

Let us restate the conditions of the theorem to fit into the ring
representation. Since for all $i=1,2,\ldots,m$, $|\Phi_i|\leq m$,
$\sa_{i-1}\cup \sa_i \subseteq \Phi_i$, and $\sa_{i-1} \cap \sa_i
= \emptyset$, then it follows that $|\sa_{i-1}| + |\sa_i| \leq m$.
Also, since $\Phi_i \cap \Phi_{i+1} \neq \emptyset$, then
$|\sa_i|\geq 1$ must hold for all $i=1,2,\ldots,m$. Finally, we
have $\sum_{i=1}^{m}{|\sa_i|}\leq m{\cal b}\frac{m}{2}{\cal c}$
(stated by theorem condition). Our task is then to provide a
proper m-coloring to all the vertices in the ring under these
stated conditions. We will consider and prove for the extreme
cases of rings whose $\sum_{i=1}^{m}{|\sa_i|}$ is equal to $m{\cal
b}\frac{m}{2}{\cal c}$. It is clear that any other instance for
which $\sum_{i=1}^{m}{|\sa_i|} < m{\cal b}\frac{m}{2}{\cal c}$ can
also be m-colored once we provide a proper m-coloring for the
extreme cases. Hence, from now on, a ring will refer to one of
these extreme instances, and is formally defined as follows.

\begin{definition} A ring $\ring$ is a
sequence $(\sa_1,\sa_2,\ldots,\sa_m)$ of $m$ cliques such that,
for all $i,j=1,2,\ldots,m$,
\begin{enumerate}
    \item $\sa_i \cap \sa_j = \emptyset$ for $i\neq j$;
    \item $\sa_i \cup \sa_{i+1}$ is a maximum clique;
    \item $|\sa_i| \geq 1$;
    \item $|\sa_i| + |\sa_{i+1}| \leq m$;
    \item $\sum_{i=1}^{m}{|\sa_i|} =  m{\cal b}\frac{m}{2}{\cal c}$.
\end{enumerate}
\end{definition}

\begin{definition}
A ring $\ring = (\sa_1,\sa_2,\ldots,\sa_m)$ is balanced (and hence is called a balanced ring) when all sets
$\sa_i$ are of the same cardinality; i.e., $|\sa_i|={\cal
b}\frac{m}{2}{\cal c}$ for $i=1,2,\ldots,m$. A ring is called unbalanced when it is not balanced.
\end{definition}

\begin{lemma}\label{claim:0}
A ring $\ring=(\sa_1,\sa_2,\ldots,\sa_m)$ is unbalanced if and only if for some $i=1,2,\ldots,m$, $|\sa_{i}|<{\cal
b}\frac{m}{2}{\cal c}$ and $|\sa_{i-1}| + |\sa_i| \leq 2{\cal
b}\frac{m}{2}{\cal c}$.
\end{lemma}

\proof The "if" part follows from the definition of the
balanced ring. Now we show the "only if" part. Let ${\cal b}\frac{m}{2}{\cal c}=n$. Since $\ring$ is not balanced, there must exist at least one $i$ for which $|\sa_i|<n$. Let
$\saa=\{i:|\sa_i|<n, i=1,2,\ldots,m\}$ and $k=|\saa|>0$. We now
show that for some $i\in\saa$, $|\sa_{i-1}| + |\sa_i| \leq 2n$.
Suppose that for all $i\in\saa$, $|\sa_{i-1}| + |\sa_i|
> 2n$. Let $\saa'=\{i-1:i\in\saa\}$ and
$\bar\saa=\{1,2,\ldots,m\}-\{\saa \cup \saa'\}$. Note that for all
$i\in\saa'$, $|\sa_i|>n$ (i.e., $i\notin\saa$) since $i+1\in\saa$
and thus $|\sa_{i+1}|<n$ and $|\sa_i| + |\sa_{i+1}| > 2n$. Hence,
$\saa$, $\saa'$ and $\bar\saa$ are disjoint. Also, note that for
all $i\in\bar\saa$, $|\sa_i|>n$. Therefore, one can write
\begin{eqnarray*}
  \sum_{i=1}^{m}{|\sa_i|} &=& \sum_{i\in\bar\saa}{|\sa_i|} +
                            \sum_{i\in\saa}{|\sa_i|} +
                            \sum_{i\in\saa'}{|\sa_i|}\\
   &=&  \sum_{i\in\bar\saa}{|\sa_i|} +
                            \sum_{i\in\saa}{(|\sa_{i-1}|+|\sa_i|)} \\
   &>& |\bar\saa|\times n + |\saa|\times 2n \\
   &=& (m-2k)\times n + k\times 2n \\
   &=& mn \\
   &=&m{\cal b}\frac{m}{2}{\cal c}
\end{eqnarray*}
which contradicts the fact that $\ring$ is a ring. \darkbox

The difficulty of the "if" part of the proof of Theorem~\ref{theorem:1} is
when $m$ is odd. When $m$ is even, the proof is relatively simple.
Therefore, we consider the two parity cases of $m$ separately.

\subsubsection{{\sc Case 1:} $m=2n+1$ is odd}
~\\
\begin{definition}
We say that a ring $\ring'=(\sa'_1,\sa'_2,\ldots,\sa'_m)$ is a
transformation of a ring $\ring=(\sa_1,\sa_2,\ldots,\sa_m)$ if
there exists a unique $i \in \{1,2,\ldots,m\}$ such that $|\sa'_i| =
|\sa_i| -1$ and either $|\sa'_{i+1}|=|\sa_{i+1}| + 1$ or
$|\sa'_{i-1}|=|\sa_{i-1}| + 1$. We then write
$\ring'=\tran_i(\ring)$.
\end{definition}

\begin{lemma}\label{claim:1}
Let $\ring=(\sa_1,\sa_2,\ldots,\sa_m)$
and $\ring'=(\sa'_1,\sa'_2,\ldots,\sa'_m)$. If $\ring'=\tran_i(\ring)$,
then $|\sa_i|\geq 2$ and either $|\sa_{i+1}| + |\sa_{i+2}| \leq
m-1$ or $|\sa_{i-1}| + |\sa_{i-2}| \leq m-1$.
\end{lemma}

\proof If $|\sa_i|< 2$, then $|\sa'_i|< 1$. Also, if
$|\sa_{i+1}| + |\sa_{i+2}| > m-1$ and $|\sa_{i-1}| + |\sa_{i-2}| >
m-1$, then $|\sa'_{i+1}| + |\sa'_{i+2}| > m$ or $|\sa'_{i-1}| +
|\sa'_{i-2}| > m$. None of the above can be true
since $\ring'$ is a ring.  \darkbox

\begin{lemma}\label{claim:2}
$\ring'$ is a transformation of $\ring$ iff $\ring$ is
a transformation of $\ring'$.
\end{lemma}

\proof If $\ring'=\tran_i(\ring)$, then it follows that
$\ring=\tran_{i-1}(\ring')$ or $\ring=\tran_{i+1}(\ring')$. \darkbox

\begin{proposition}\label{lemma:1}
For every unbalanced ring $\ring$, there exists a sequence
of a finite number $k$ of rings $(\ring^1,\ring^2,\ldots\ring^k)$
such that $\ring^1$ is a transformation of the balanced ring
$\ring^0$, $\ring^i$ is a transformation of $\ring^{i-1}$ for
$i=2,3,\ldots,k$, and $\ring$ is a
transformation of $\ring^k$.
\end{proposition}

\proof Let $\ring^0=(\sa^0_1,\sa^0_2,\ldots,\sa^0_m)$ be the
balanced ring and $\ring\neq \ring^0$ be a ring. Instead of
showing that $\ring$ can be obtained via a finite number of
transformations from $\ring^0$, we show the opposite; that is, we show that
$\ring^0$ can be obtained through a finite number of
transformations from $\ring$. The proof will then follow from
{\sc Lemma}~\ref{claim:2}. We transform $\ring$ to $\ring^0$ by moving
vertices across the sets $\sa_i$ until we obtain
$\sa_i=n$ for all $i$ (which results in $\ring^0$). The procedure
takes place in a finite number of iterations, each of which
involves a finite number of transformations. The transformation
procedure is as follows.
\begin{enumerate}
\item \label{one} Make $\saa=\{i : |\sa_{i}| < n,
i=1,2,\ldots,m\}$,

\item \label{iter1:1} If $\saa = \emptyset$ (i.e., ring is
balanced), then stop. Else, pick any $i\in\saa$ satisfying
{\sc Lemma}~\ref{claim:0},

\item \label{iter1:2} Find any $j>i$ such that $|\sa_j|>n$, and
for all $j'=i+1,i+2,\ldots,j-1$, $|\sa_{j'}| \leq n$ (it is easy
to argue that there exists one since $\ring$ is not balanced),

\item Move a vertex from $\sa_{j}$ to $\sa_i$. This iteration is
attainable in $j-i$ transformations. That is, the first vertex to
be moved is from the set $\sa_{j}$ to the set $\sa_{j-1}$, then
from $\sa_{j-1}$ to $\sa_{j-2}$, $\dots$, from $\sa_{i+1}$ to
$\sa_{i}$. Go back to Iteration~\ref{one}.
\end{enumerate}
There are two points that are worth mentioning. First, the choice of $j$
imposed through Iteration~\ref{iter1:2} assures that when the
vertex is moving down, the resulted rings are transformations of
their previous ones. This is because the property $|\sa_i| +
|\sa_{i+1}| \leq 2n+1$ is not violated when moving a vertex from
$\sa_{i+2}$ down to $\sa_{i+1}$. This property is also assured for
the last move (from $\sa_{i+1}$ to $\sa_i$) by the choice of $i$
made at Iteration~\ref{iter1:1}.
Second, it is important to notice that the above procedure
terminates due to the fact that, by the end of each four
iterations, the number $\sum_{i\in\saa}{(n-|\sa_i|)}$ decreases by
one. Therefore, this number will eventually go to zero resulting in $\saa$
being empty; i.e., ring is balanced.
\darkbox

Let $\ring=(\sa_1,\sa_2,\ldots,\sa_m)$ be any given ring. For every
$i=1,2,\ldots,m$, let $\Pi_i$ denote the set of all sets that each
contains exactly one (any one) vertex of each of the sets $\sa_i,
\sa_{i+2},\sa_{i+4},\ldots,\sa_{m+i-5},\sa_{m+i-3}$. Formally,
$\Pi_i=\{\{j_0,j_1,\ldots,j_{n-1}\}\;|\;j_k\in\sa_{i+2k},
k=0,1,\ldots,n-1\}$. That is, any set of the form
$\{i,i+2,i+4,\ldots,m+i-5,m+i-3\}$ is in $\Pi_i$, where an index
$j$ in the set refers to any vertex in $\sa_j$. Observe that each
set in $\Pi_i$ is an independent set since no two vertices in it
present an edge. Because in a cycle of $2n+1$ vertices at most $n$
vertices can be chosen such that no two vertices share an edge
between them, an independent set of $\ring$ can at most contain
$n$ vertices. Hence, for all $i$, each set in $\Pi_i$ is a maximum
independent set (since each contains $n$ vertices).
Fig~\ref{fig:diagram} shows a graphical representation of
$\Pi_1,\Pi_2,\ldots,\Pi_m$ as a Diagram. An index $j$ in $\Pi_i$
refers to any vertex of $\sa_j$. An arrow linking two sets $\Pi_i$
and $\Pi_j$ indicates that any maximum independent set in $\Pi_i$
can be transformed to a maximum independent set in $\Pi_j$ by
substituting a vertex for another. Vertices that can be
substituted are indicated above the arrow.
\begin{figure}
  \centerline{
  \includegraphics[width=0.85\columnwidth]{diagram.pstex}}
  \caption{Coloring Diagram.}\label{fig:diagram}
\end{figure}

\begin{proposition}\label{claim:3}
$\Pi_1,\Pi_2,\ldots,\Pi_m$ are pairwise disjoint.
\end{proposition}

\proof Note that each maximum independent set in $\Pi_i$ is
missing the two consecutive vertices $m+i-2$ and $m+i-1$, while
exactly one of these two vertices ($m+i-2$ or $m+i-1$) is present
in every other maximum independent set in $\Pi_j$ for $j\neq i$.
Hence each set in $\Pi_i$ does not belong to any set $\Pi_j$ for
$j\neq i$.\darkbox

\begin{lemma}\label{claim:4}
Every maximum independent set of any ring belongs to one $\Pi_i$
for some $i=1,2,\ldots,m$.
\end{lemma}

\proof Since each maximum independent set must not contain
two vertices belonging to two consecutive sets $\sa_i$ and
$\sa_{i+1}$ for some $i$ and there are only $m$ possible different
pairs of consecutive vertices, then each maximum independent set
must not contain one of these $m$ pairs. Hence, every maximum
independent set must belong to one of the $m$ sets
$\Pi_1,\Pi_2,\ldots,\Pi_m$.
\darkbox

\begin{proposition}\label{claim:5}
A ring is balanced if and only if its vertices can be partitioned
into exactly $m$ disjoint maximum independent
sets $\pi_1^0,\pi_2^0,\ldots,\pi_m^0$ such that $\pi_i^0 \in \Pi_i$ for all $i=1,2,\ldots,m$.
\end{proposition}

\proof Let $\ring=(\sa_1,\sa_2,\ldots,\sa_m)$ be the
balanced ring; i.e, $|\sa_i|=n$ for all $i$. Let
$\{v_{i,1},v_{i,2},\ldots,v_{i,n}\}$ denote the vertices in
$\sa_i$ for all $i$. For each $i$, we define the set $\pi^0_i$ to
be $\{v_{i,1},v_{i+2,2},v_{i+4,3},\ldots,v_{m+i-3,n}\}$. For
example, $\pi^0_1=\{v_{1,1},v_{3,2},v_{5,3},\ldots,v_{m-2,n}\}$,
$\pi^0_2=\{v_{2,1},v_{4,2},v_{6,3},\ldots,v_{m-1,n}\}$, and so
forth. Clearly, $\pi^0_i\in \Pi_i$ for all $i=1,2,\ldots,m$ and
hence $\pi_1^0,\pi_2^0,\ldots,\pi_m^0$ are $m$ disjoint maximum
independent sets (follows from {\sc Proposition}~\ref{claim:3}). Also, one can
easily see that the vertices of each set $\sa_i$ are contained in
the $n$ distinct sets $\pi^0_i,\pi^0_{i+2},\ldots,\pi^0_{m+i-3}$
(each vertex is contained in a different set). Hence all the
vertices of the ring are partitioned into the $m$ disjoint maximum
independent sets. Now suppose that the vertices of a ring can be
split into $m$ disjoint maximum independent sets each of which
belongs to a different $\Pi_i$ for some $i$. Then, all the
vertices of each set $\sa_i$ are contained in exactly $n$
independent sets, each of which contains no more than one vertex.
Hence, $|\sa_i|=n$ for all $i$; i.e., the ring is balanced.
\darkbox

\begin{lemma}\label{claim:6}
A ring has exactly $m$ disjoint maximum independent sets if and only if it is m-colorable.
\end{lemma}

\proof The forward direction is trivial. Since there are $m$
disjoint sets each of which has $n$ vertices (since they are
maximum), then at least $mn$ different vertices can be properly
m-colored. This can be done by coloring each independent set with
a different color. This results in a proper m-coloring since a
ring has $mn$ vertices. Now suppose that a given ring is
m-colorable. Then, each color must have been used by at most $n$
vertices because of the fact that in a cycle of $2n+1$ vertices at
most $n$ vertices can be chosen such that there is no edge between
any two of them. Now since a ring has $mn$ vertices and there are
$m$ colors, each color must have been used by exactly $n$
vertices. Therefore, there must be at least $m$ disjoint
independent sets of $n$ vertices each. Hence there must be at
least $m$ disjoint maximum independent sets (since size of these
independent sets is $n$). Again because a ring has $mn$ vertices,
there must be exactly $m$ disjoint maximum independent sets.
\darkbox

\begin{lemma}\label{claim:7}
Every set in $\Pi_i$ for $i=1,2,\ldots,m$ can be transformed
to a set in $\Pi_{i+2}$ by substituting $i$ for ${i-1}$ or to a
set in $\Pi_{i-2}$ by substituting $i-3$ for $i-2$.
\end{lemma}

\proof Note that each set in $\Pi_i$ does not contain a
vertex $i-1$, nor a vertex $i-2$, but contains a vertex $i$.
Hence, if $i$ is substituted for $i-1$, then $i-1$ still does
share edges with any of the vertices in the set. This new obtained
set is indeed in $\Pi_{i+2}$. Similarly, we can prove for the case
of $\Pi_{i-2}$.
 \darkbox


Note that the above two transformations are the only two
transformations involving the substitution of only one vertex that
transform a set in $\Pi_i$ to another in $\Pi_j$'s. This is because
the insertion of any other vertex will interfere with an already
existing vertex. The two-way arrows in the Coloring Diagram
provided in Fig.~\ref{fig:diagram} shows all possible
substitutions. This diagram will be an essential part of the
proposed coloring algorithm that we describe in the next section.

\begin{proposition}\label{lemma:2}
Let $\ring$ and $\ring'$ be two rings that are transformations of one another.
$\ring$ is m-colorable if and only if $\ring'$ is m-colorable.
\end{proposition}

\proof We only need to prove one direction; the
other follows from {\sc Lemma}~\ref{claim:2}. Suppose
$\ring'=(\sa'_1,\sa'_2,\ldots,\sa'_m)$ is a transformation of
$\ring=(\sa_1,\sa_2,\ldots,\sa_m)$ and $\ring$ is m-colorable.
Also, let $p$ be the index such that $|\sa'_p| = |\sa_p|
- 1$ and $|\sa'_{p+1}| =|\sa_{p+1}| + 1$ (the case where
$|\sa'_{p-1}| =|\sa_{p-1}| + 1$ can be proven similarly). We need
to show that $\ring'$ is m-colorable. From {\sc Lemma}~\ref{claim:1}, it
follows that $|\sa_{p+1}| + |\sa_{p+2}| \leq m-1$ must hold.
First, observe that one and only one vertex belonging to either
one the two successive sets, $\sa_{p+1}$ and $\sa_{p+2}$, belongs
to each set in $\Pi_i$, except those in $\Pi_{p+3}$ which contain
none of the two vertices. Second, since $\ring$ is m-colorable,
then it follows from {\sc Lemma}~\ref{claim:6} that the vertices of
$\ring$ are the union of $m$ disjoint maximum independent sets.
Now {\sc Lemma}~\ref{claim:4} implies that each of these $m$ sets must
belong to one $\Pi_i$ for $i=1,2,\ldots,m$. Hence, in order for
$|\sa_{p+1}| + |\sa_{p+2}|$ to be less than or equal to $m-1$,
$\Pi_{p+3}$ must be one of these $m$ disjoint sets. Now using
{\sc Lemma}~\ref{claim:7}, one can transform any maximal independent set
in $\Pi_{p+3}$ to the maximal independent set in $\Pi_{p+1}$ by
substituting the vertex $p$ for $p+1$. Hence, the new $m$ maximum
independent sets whose union is $\ring'$ are also disjoint.
$\ring'$ is m-colorable by {\sc Lemma}~\ref{claim:6}.
\darkbox ~\\

Now we provide the proof of the "if" part of the main theorem ({\sc Theorem}~\ref{theorem:1}) when $m$
is odd, which now follows from the derived propositions. First, using
{\sc Proposition}~\ref{claim:5}, one can properly color the balanced ring with
m different colors (each maximal independent set is colored with a
different color). Hence the balanced ring is m-colorable. Second,
by {\sc Proposition}~\ref{lemma:1} we know that any ring can be obtained from
the balanced ring through a finite number of transformations.
Finally, a ring has a proper $m$ coloring follows from
{\sc Proposition}~\ref{lemma:2}. This ends the proof of sufficiency part of
the {\sc Theorem}~\ref{theorem:1} when $m$ is odd. Note that this proof is nothing but a coloring algorithm that colors $\phi$ with exactly m colors. This algorithm is formally presented in Section~\ref{sec:algorithm}.

\subsubsection{{\sc Case 2:} $m=2n$ is even} When $m$ is
even, the proof (and hence a coloring algorithm) is simple; i.e.,
it does not require rearrangement of colors. Let
$\ring=(\sa_1,\sa_2,\ldots,\sa_m)$ be a ring such that
$\sum_{i=1}^{m}{\sa_i} = mn$. The following is a proper m-coloring
of $\ring$. Let $\{c_1,c_2,\ldots,c_m\}$ denote the set of m
different colors. For each $i=1,2,\ldots,m$, color the vertices in
$\sa_i$ with $\{c_1,c_2,\ldots,c_{|\sa_i|}\}$ if $i$ is odd and
with $\{c_{m},c_{m-1},\ldots,c_{m-|\sa_{i}|+1}\}$ if $i$ is even.
Clearly, this is a proper m-coloring because, for any $i$, none of
the vertices in $\sa_i$ share the same color with a vertex in
$\sa_{i-1} \cup \sa_{i+1}$ (note that these vertices are the only
ones that share edges with vertices in $\sa_i$). This is due to
the fact that $|\sa_{i-1}| + |\sa_{i}| \leq m$ and $|\sa_{i}| +
|\sa_{i+1}| \leq m$.

\subsection{The "only if" part proof}
Let $\ring=(\sa_1,\sa_2,\ldots,\sa_m)$ be an m-colorable ring.
Since $\ring$ is m-colorable, then there must exist at most $m$
disjoint independent sets whose union is $\ring$. Now because each
independent set cannot contain more than ${\cal b}\frac{m}{2}
{\cal c}$ vertices (due to cycles of length $m$), then at most the
ring contains $m{\cal b} \frac{m}{2} {\cal c}$ vertices.

\section{A Coloring Algorithm}
\label{sec:algorithm}
The proposed coloring algorithm that we present in this section follows from the proof presented in Section~\ref{sec:theorem}.
After describing the algorithm for the
general case, we illustrate it through an example with $m=7$. Let
us consider a ring $\ring=(\sa_1,\sa_2,\ldots,\sa_m)$ and let
$m=2n+1$. The key idea of the proposed coloring algorithm is to
partition the vertices of $\ring$ into $m$ disjoint maximum
independent sets---hence, coloring each of them with a different
color yields to a proper m-coloring of $\ring$. In
Section~\ref{sec:theorem}, we showed that $\ring$ is indeed
partitionable into $m$ disjoint maximum independent sets, each of
which must belong to $\Pi_i$ for some $i=1,2,\ldots,m$. Let
$\si_i$ denote the number of maximum independent sets among these
$m$ sets which belong to $\Pi_i$ (i.e., $\sum^{m}_{i=1}{\si_i}=m$). The coloring algorithm consists then
of determining $\si_i$ for all $i$. We now describe
the different steps of the algorithm.

\begin{enumerate}
\item \label{step:1} $\si_i\leftarrow 1$ for all $i=1,2,\ldots,m$,

\item $\saa \leftarrow \{ i : |\sa_i| < n, i=1,2,\ldots,m\}$,

\item While $\saa \neq \emptyset$, do
\begin{enumerate}
    \item
    Pick any $i\in \saa$ such that $|\sa_{i-1}| + |\sa_i| \leq 2n$,

    \item Pick any $j>i$ such that $|\sa_j|>n$ and $|\sa_{j'}| \leq n$
for all $j'=i+1,i+2,\ldots,j-1$,

    \item \label{step:3c}
        $\si_{k} \leftarrow \si_{k} - 1$ for $k=j+2,j+1$,
    \item \label{step:3d}
        $\si_{k} \leftarrow \si_{k} + 1$ for $k=i+2,i+1$,
    \item $|\sa_{j}| \leftarrow |\sa_{j}| - 1$,

    \item $|\sa_{i}| \leftarrow |\sa_{i}| + 1$,

    \item $\saa \leftarrow \{ i : |\sa_i| < n, i=1,2,\ldots,m\}$.

\end{enumerate}
\end{enumerate}
Although the above coloring algorithm follows straightly from the
proof the theorem provided in Section~\ref{sec:theorem}, it is
worth bringing the attention to the following three points. First, note
that instead of transforming the balanced ring to the ring in
question, we proceed in the opposite direction; that is, we apply a finite number of transformations to the unbalanced ring until the balanced ring is obtained.
Second,
Step~\ref{step:1}) follows from the fact that $\si_i=1$ for all
$i=1,2,\ldots,m$ when the ring is balanced. Third, Steps~\ref{step:3c})
and~\ref{step:3d}) follow from {\sc Lemma}~\ref{claim:7} which states
that moving a vertex from $\sa_k$ to $\sa_{k-1}$ is equivalent to
substituting a maximum independent set in $\Pi_k$ for one in
$\Pi_{k+2}$ for all $k=j,j-1,\ldots,i+1$ (the in the lemma, we prove
that such a substitution exists).

\begin{figure}
  \centerline{
  \includegraphics[width=0.75\columnwidth]{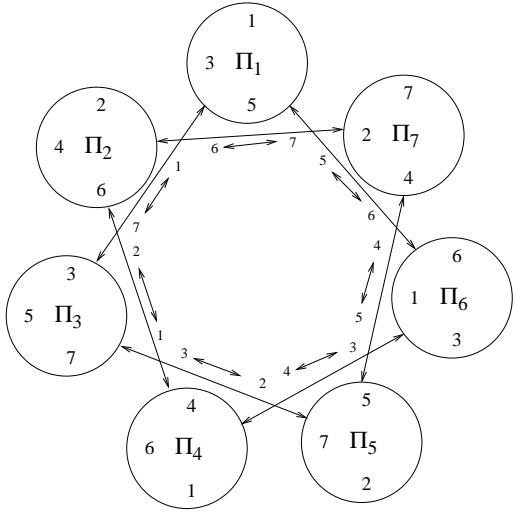}}
  \caption{Coloring Diagram for $m=7$.}\label{fig:exp}
\end{figure}

{\bf \sc Example.}
Let us now apply the coloring algorithm to an example. Consider
the case when $m=7$ (i.e., $n=3$). The coloring diagram for $m=7$ is shown in
Fig.~\ref{fig:exp}.
Let $\ring$ be the example ring $(\sa_1,\sa_2,\ldots,\sa_7)$ with
$|\sa_1|=5$, $|\sa_2|=2$, $|\sa_3|=3$, $|\sa_4|=4$, $|\sa_5|=1$,
$|\sa_6|=4$, and $|\sa_7|=2$. All the steps of the coloring algorithm are executed and shown in Table~\ref{tab:exp}. When the algorithm terminates, the $m$ disjoint maximum independent sets of $\ring$ are all provided through the numbers $\si_i$ for all $i=1,2,\ldots,m$, which are the outcome of the algorithm. As shown in Table~\ref{tab:exp}, these numbers are $\si_2=\si_3=\si_5=0$, $\si_1=1$, and $\si_4=\si_6=\si_7=2$.
\begin{table}
\small
  \centering
  \caption{Coloring algorithm steps for the example}\label{tab:exp}
  \begin{tabular}{||c||c|c|c|c|c||} \hline
    Iterations & initial & Iter. $1$ & Iter. $2$ & Iter. $3$ & Iter. $4$ \\ \hline
    $|\sa_1|$ & $5$ & $4$ & $4$ & $4$ & $3$ \\ \hline
    $|\sa_2|$ & $2$ & $2$ & $3$ & $3$ & $3$ \\ \hline
    $|\sa_3|$ & $3$ & $3$ & $3$ & $3$ & $3$ \\ \hline
    $|\sa_4|$ & $4$ & $4$ & $3$ & $3$ & $3$ \\ \hline
    $|\sa_5|$ & $1$ & $1$ & $1$ & $2$ & $3$ \\ \hline
    $|\sa_6|$ & $4$ & $4$ & $4$ & $3$ & $3$ \\ \hline
    $|\sa_7|$ & $2$ & $3$ & $3$ & $3$ & $3$ \\ \hline \hline
    $\si_1$ & $1$ & $2$ & $2$ & $1$ & $1$ \\ \hline
    $\si_2$ & $1$ & $1$ & $1$ & $1$ & $0$ \\ \hline
    $\si_3$ & $1$ & $0$ & $1$ & $1$ & $0$ \\ \hline
    $\si_4$ & $1$ & $1$ & $2$ & $2$ & $2$ \\ \hline
    $\si_5$ & $1$ & $1$ & $0$ & $0$ & $0$ \\ \hline
    $\si_6$ & $1$ & $1$ & $0$ & $1$ & $2$ \\ \hline
    $\si_7$ & $1$ & $1$ & $1$ & $1$ & $2$ \\ \hline \hline
    $\saa$ & $\{2,5,7\}$ & $\{2,5\}$ & $\{5\}$ & $\{5\}$ & $\emptyset$ \\ \hline
    $i$ & $7$ & $2$ & $5$ & $5$ & $-$ \\ \hline
    $j$ & $1$ & $4$ & $6$ & $1$ & $-$ \\ \hline \hline
    \end{tabular}
\end{table}
Therefore, the vertices of $\ring$ can be partitioned into the
following $7$ disjoint maximum independent sets: one set of the form
$\{1,3,5\}$ ($\si_1=1$); two sets each of the form $\{4,6,1\}$ ($\si_4=2$);
two sets each of the form $\{6,1,3\}$ ($\si_6=2$); and two sets each of the form
$\{7,2,4\}$ ($\si_7=2$). This yields to $|\sa_1|=5$, $|\sa_2|=2$,
$|\sa_3|=3$, $|\sa_4|=4$, $|\sa_5|=1$, $|\sa_6|=4$, and
$|\sa_7|=2$. Coloring each maximum independent set of the $7$ sets
with a different color results in a proper $7$-coloring of
$\ring$.

\section{Conclusion}
\label{sec:conc}
In this work, we derived and proved a sufficient and necessary condition on m-colorability of odd m-clique holes, and proposed a coloring algorithm that colors m-clique holes with exactly m colors. 

\bibliography{main}
\bibliographystyle{plain}
\end{document}